\documentclass[10pt,amsfonts,amssymb,amsmath,aps,prl,twocolumn,showpacs,reprint]{revtex4-1}

\usepackage{graphicx}  
\usepackage{dcolumn}   
\usepackage{bm}        

\expandafter\ifx\csname package@font\endcsname\relax\else
 \expandafter\expandafter
 \expandafter\usepackage
 \expandafter\expandafter
 \expandafter{\csname package@font\endcsname}%
\fi

\usepackage{xcolor}

\renewcommand{\vec}[1]{{\ensuremath{\bm{\mathrm{#1}}}}}
\usepackage{dsfont}

\newcommand{\E}[1]{\ensuremath{{10^{#1}}}}

\begin{document}

\title{Higher order exchange interactions leading to metamagnetism in FeRh}

\author{Joseph Barker}
\email{joseph.barker@imr.tohoku.ac.jp}
\author{Roy W. Chantrell}
\affiliation{Department of Physics, University of York, York YO10 5DD, United Kingdom}

\begin{abstract}
  The origin of the metamagnetic antiferromagnetic-ferromagnetic phase transition of FeRh is a subject of much debate. Competing explanations invoke magnetovolume effects and purely thermodynamic transitions within the spin system. It is experimentally difficult to observe the changes in the magnetic system and the lattice simultaneously, leading to differing conclusions over which mechanism is responsible for the phase transition. A non-collinear electronic structure study by Mryasov [O.N. Mryasov, Phase Transitions \textbf{78}, 197 (2005)] showed that non-linear behavior of the Rh moment leads to higher order exchange terms in FeRh. Using atomistic spin dynamics (ASD) we demonstrate that the phase transition can occur due to the competition between bilinear and the higher order four spin exchange terms in an effective spin Hamiltonian. The phase transition we see is of first order and shows thermal hysteresis in agreement with experimental observations. Simulating sub-picosecond laser heating we show an agreement with pump-probe experiments with a ferromagnetic response on a picosecond timescale.
\end{abstract}

\maketitle

The metamagnetic transformation of FeRh from antiferromagnetic (AFM) to ferromagnetic (FM) ordering has been known for over 70 years~\cite{Fallot:1939ta}. This dramatic transformation occurs at a temperature of $T_{\mathrm{M}}=350$K, although changes in composition~\cite{Staunton:2014vx}, strain~\cite{Cherifi:2014uh}, doping~\cite{Thiele:2003hf} and magnetic fields~\cite{Annaorazov:1992tl} can move the transition temperature significantly. The accessibility and tunability of the transition, as well as the different magnetic behavior that AFM and FM order provide, means that FeRh could be used for some interesting technological applications~\cite{Annaorazov:1992tl,Annaorazov:2002vk,Thiele:2003hf,Guslienko:2004fh}.

Below the transition temperature, FeRh exists as an antiferromagnet where the Fe site has moment $|\vec{m}_{\mathrm{Fe}}| \simeq 3.15\mu_{\mathrm{B}}$ and the Rh site has no net magnetic moment. Above the transition temperature the Fe moments re-align ferromagnetically and the Rh site forms a moment of $|\vec{m}_{\mathrm{Rh}}| \simeq 1.00\mu_{\mathrm{B}}$ while the Fe moment is largely unchanged. There is also a 1\% expansion of the unit cell volume in the FM phase. Debate exists about the driving force behind the phase transition. The contention concerns whether the expansion of the unit cell through the phase transition alters the magnetic state, or whether a thermodynamic phase transition in the magnetic state drives the lattice expansion. As yet neither experiments nor theory have been conclusive on this matter.

Non-collinear electronic structure studies have shown a non-linear dependence of the direction and magnitude of the Rh moment on the Weiss field from the surrounding Fe moments~\cite{Mryasov:2005cda}. This unusual behavior allows one to write an effective spin Hamiltonian which contains only the Fe degrees of freedom where the non-linear induced Rh moment leads to higher order effective exchange contributions of biquadratic and four spin order. It has been suggested that the competition between exchange interactions of different orders around the transition temperature could drive the phase transition, with the volume expansion occurring as a subsidiary effect, thus explaining observations of sub-picosecond laser heating where the reponse of the magnetic system was demonstrated to respond faster than that of the lattice~\cite{Ju:2004vp}.

In this Letter we demonstrate that it is the thermally driven competition between the bilinear and four spin contributions to the effective Fe-Rh-Fe exchange which lead to the AFM-FM phase transition. Specifically, it will be shown that the transition is a direct result of the differential thermal scaling of the two terms. Using a minimal set of interaction parameters we are able to reproduce the experimentally measured temperature dependent magnetization of the ferromagnetic phase. The phase transition of this model is of first order, in agreement with experimental observations and this leads to thermal hysteresis at the transition. Importantly, our use of the dynamical approach of atomistic spin dynamics (ASD) allows the study of the time scale on which the phase transition occurs during sub-picosecond laser heating.  We find that it is possible for FM order to form on a picosecond time scale, but only if the intrinsic Gilbert damping is large, of the order of $\alpha = 0.1$. As this model includes no lattice degrees of freedom, it demonstrates that it is possible for the phase transition to occur only from the thermodynamic behavior of the magnetic interactions within the spin system, in agreement with other theoretical calculations~\cite{Gu:2005us}.

The Hamiltonian for FeRh proposed by Mryasov based on first principles calculations~\cite{Mryasov:2005cda} contains only Fe degrees of freedom. The effect of the Rh site enters this Hamiltonian by an effective Fe-Rh-Fe exchange term. In principle the Hamiltonian is similar to that of FePt which has been successfully modeled previously~\cite{Mryasov:2007kj,Kazantseva:2008wc,Barker:2010hf}. The linear dependence of the size of the Pt moment on the Weiss field from the Fe moments caused an enhancement of the ferromagnetic bilinear exchange energy between Fe moments. In the case of FeRh, the quadratic dependence of the Rh moment on the Weiss field leads to higher order effective exchange terms. It is the competition between ferromagnetic bilinear terms and antiferromagnetic, biquadratic terms which has been suggested as a mechanism for the phase transition previously~\cite{Ju:2004vp}.

An important detail which is somewhat unclear in Ref.~\onlinecite{Mryasov:2005cda}, is that the effective Hamiltonian which is written must be fully expanded to also include four spin exchange terms $(\vec{S}_{i}\cdot\vec{S}_{j})(\vec{S}_{k}\cdot\vec{S}_{l})$, as well as the biquadratic terms $(\vec{S}_{i}\cdot\vec{S}_{j})^{2}$. Indeed, as biquadratic interactions are degenerate in energy for AFM and FM ordering, it must be the four spin terms which lead to the AFM order at low temperature if an AFM-FM phase transition is to take place. Other models and theoretical works have also found that effects of the order $m^4$ were needed to stabilize the phase transition~\cite{Derlet:2012vd} or to fit first principles results~\cite{Staunton:2014vx}.

\begin{figure}
  \includegraphics[width=0.48\textwidth]{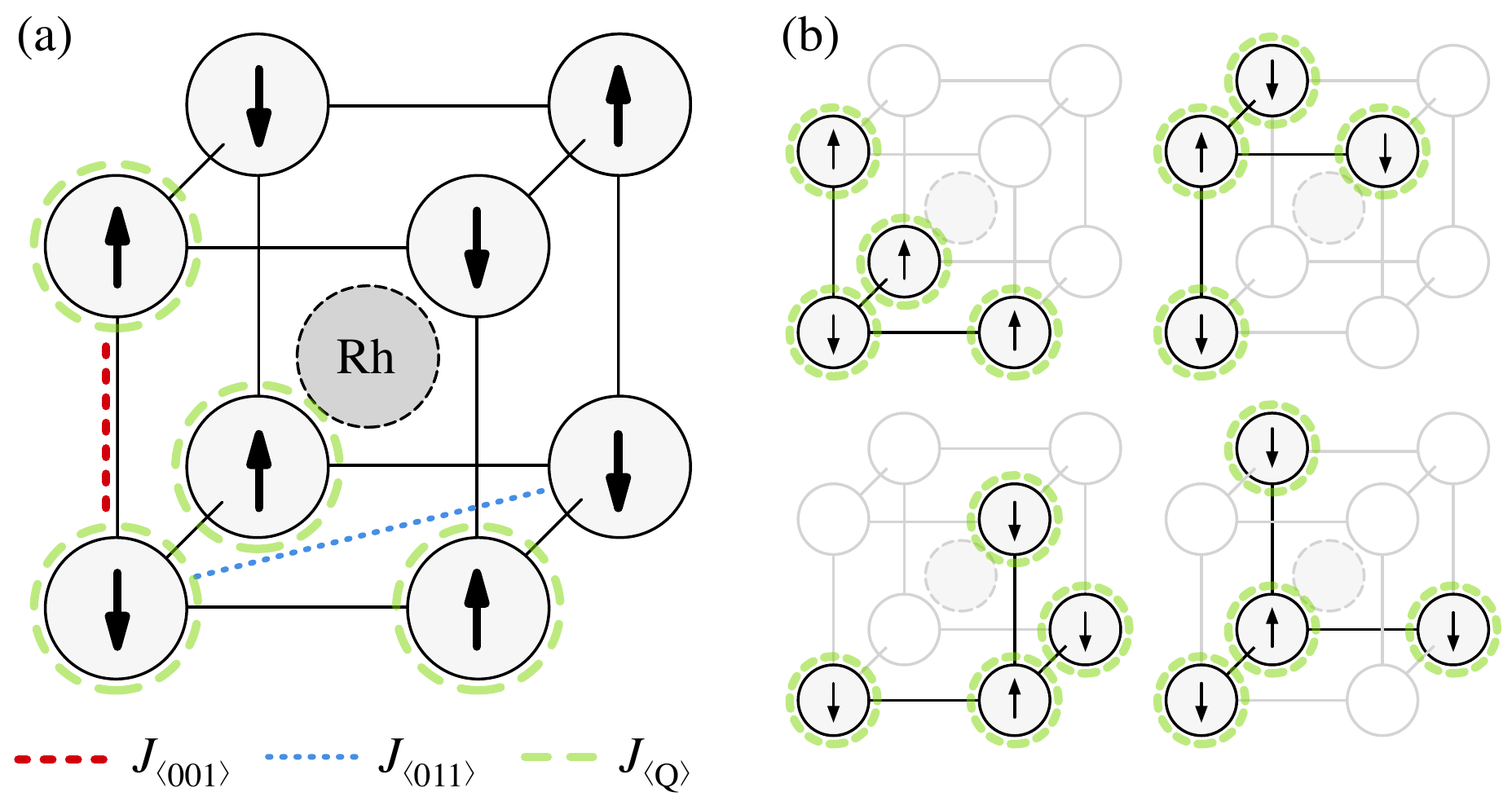}
  \caption{(Color Online)
    In this simplified model we include (i) nearest neighbor exchange interactions
    ($J_{\langle 001 \rangle}$) (ii) next nearest neighbor exchange interactions ($J_{\langle 011 \rangle}$) (iii) four spin
    interactions around the basic quartet ($D_{\langle Q \rangle}$).
  }
\label{fig:ferh_unitcell}
\end{figure}

We use a simplified Hamiltonian with the aim of qualitatively demonstrating that the AFM-FM phase transition can occur from a purely magnetic interaction Hamiltonian. The complexity of parameterizing a spin model with long-ranged bilinear, biquadratic and four spin exchange terms from first principles is prohibitive and the number of interactions contained in such a model would make an efficient implementation difficult. Thus we reduce the complex Hamiltonian in Ref.~\onlinecite{Mryasov:2005cda} to the competition between bilinear and four spin interactions
\begin{align}
\begin{split}
  \mathcal{H} = &-\sum_{i,j} J_{ij} \vec{S}_{i}\cdot\vec{S}_{j} +
            \tfrac{1}{3}\sum_{i,j,k,l} D_{ijkl} \left[ (\vec{S}_{i}\cdot\vec{S}_{j})(\vec{S}_{k}\cdot\vec{S}_{l})\right.\\
            &\left. + (\vec{S}_{i}\cdot\vec{S}_{k})(\vec{S}_{j}\cdot\vec{S}_{l}) + (\vec{S}_{i}\cdot\vec{S}_{l})(\vec{S}_{k}\cdot\vec{S}_{j})\right].
\label{eq:hamiltonian}
\end{split}
\end{align}
The spin degrees of freedom $\vec{S}_{i}$ in this Hamiltonian represent only the Fe moments on a simple cubic lattice. The Rh moments on the central site of the B2 CsCl structure are not modeled explicitly, the effect of the induced Rh moment is through the higher order coupling of the Fe moments. Bilinear nearest neighbor, $J_{\langle 001 \rangle}$, and next-nearest neighbor, $J_{\langle 011 \rangle}$, terms (Fig.~\ref{fig:ferh_unitcell}a) give the inter- and intra-sublattice exchange interactions and contributions from both Fe-Fe and the Fe-Rh-Fe interactions are combined to a single value. The simplest form for the four spin term is to include only the `basic quartets', $D_{\langle Q \rangle}$, of the simple cubic lattice (Fig.~\ref{fig:ferh_unitcell}b)~\cite{Mouritsen:1983tx}.

Despite the approximations we have made in simplifying the Hamiltonian, solving this system of interacting spin moments using a statistical approach is important for the investigation of phase transitions and is a significant advance on mean field approaches~\cite{Onyszkiewicz:1978wo,Ju:2004vp}. Our interest in the dynamical behavior and characteristic time scales of this system leads us to the use of ASD to calculate the thermodynamic and dynamic properties. Another advantage of using ASD is its greater computational efficiency in this case than Monte Carlo methods in determining equilibrium states. The four spin interaction term is computationally expensive and cannot easily be computed using Fourier based techniques. However, our GPU (graphical processing unit) accelerated ASD implementation allows the solution of the system of coupled equations 1-2 orders of magnitude faster than CPU implementations, making an over damped Langevin approach significantly faster than Monte-Carlo integration techniques and allowing for good statistical sampling on large spin lattices.

The ASD approach is based on the solution of the Landau-Lifshitz-Gilbert equation of motion with Langevin dynamics
\begin{equation}
    \frac{\partial \vec{S}_{i}}{\partial t} = -\frac{\gamma}{1+\alpha^2}\left[ \vec{S}_{i} \times
    \vec{H}_{i} + \alpha \vec{S}_{i}\times(\vec{S}_{i}\times\vec{H}_{i})\right].
\label{eq:llg}
\end{equation}
$\mu_{s}=3.15$~$\mu_{B}$ is the magnitude of the Fe magnetic moment~\cite{Moruzzi:1992vg}, the Rh moments are not explicitly modeled. $\gamma=1.024\gamma_{e}=1.80224\times10^{11}$~rad~s$^{-1}$~T$^{-1}$ is the gyromagnetic
ratio~\cite{Mancini:2013ut} and $\alpha$ is the Gilbert damping, the value of which we discuss later. For the calculation of equilibrium thermodynamic properties we use an over-damped Langevin approach with $\alpha = 1$.
The effective field on each lattice site is
\begin{equation}
  \vec{H}_{i} = -\frac{1}{\mu_{s}}\frac{\partial \mathcal{H}}{\partial \vec{S}_{i}} + \vec{\xi}_{i}
\label{eq:llg_field}
\end{equation}
and $\vec{\xi}_{i}$ is a stochastic term which represents on site thermal fluctuations in the white noise limit. One outstanding question from previous models of FeRh is whether or not the thermal fluctuations of the Rh moment play a significant role in the phase transition. In the Blume-Capel like Ising model used by Gruner et al.~\cite{Gruner:2003kl}, instability of the Rhodium moment leads to the phase transition. However, given the magnetically soft nature of FeRh, an Ising model is less than ideal for representing the spin system. In this work it is assumed that the direction and size of the Rh moment is determined solely by the Weiss field from the Fe and thermally induced longitudinal fluctations of the Rh moment are ignored. Hence we are describing a completely different mechanism for the phase transition to that of Gruner et al.

\begin{figure}
  \includegraphics[width=0.45\textwidth]{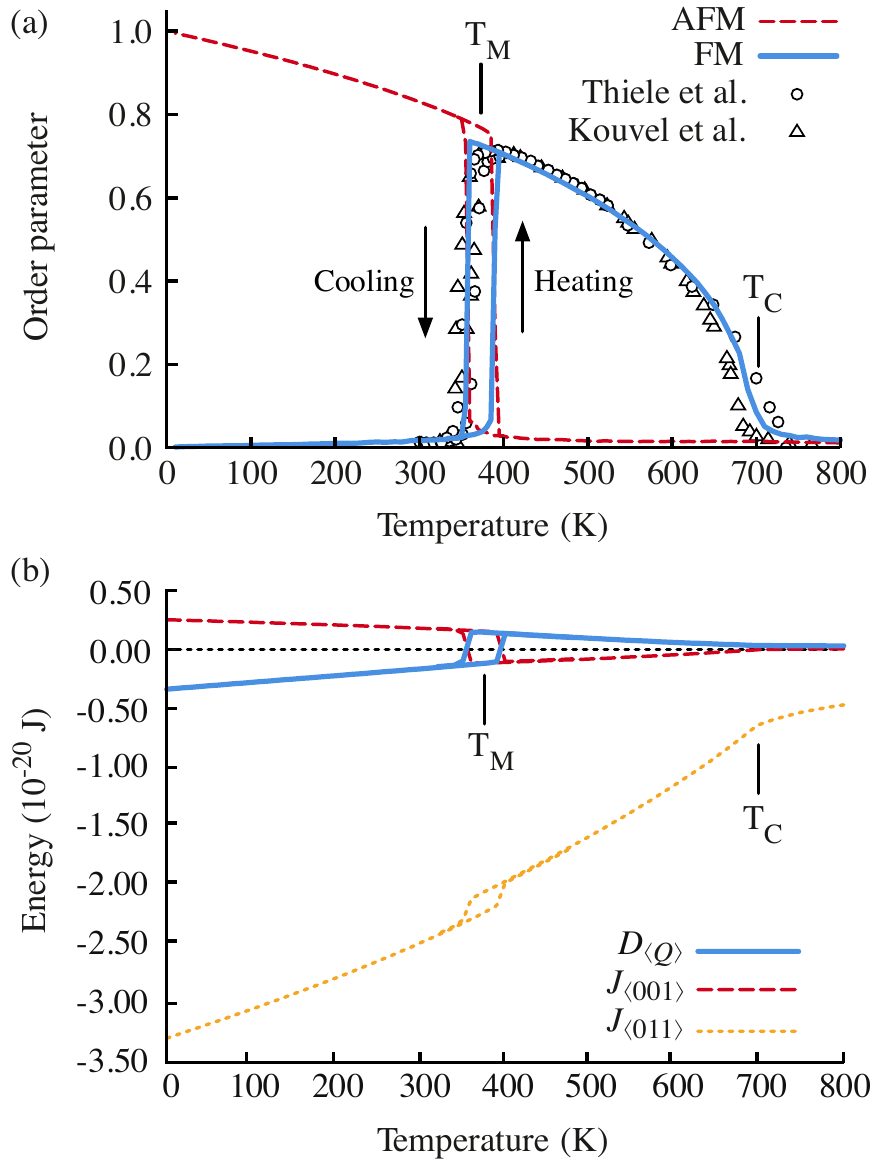}
  \caption{(Color Online)
    (a) An AFM-FM phase transition is observed in the model Hamiltonian Eq.~(\ref{eq:hamiltonian}). A good fit to experimental measurements of FeRh~\cite{Thiele:2003hf,Kouvel:1962ba} can be found by choosing the parameters $J_{\langle 001 \rangle}$, $J_{\langle 011 \rangle}$ and $D_{\langle Q \rangle}$. (b) Calculation of the energy contributions from each term in the Hamiltonian shows a cross over dependence at $T_{\mathrm{M}}$ due to the larger thermal scaling exponent of the four spin term than the FM bilinear term $J_{\langle 001 \rangle}$ with which it competes.
  }
\label{fig:ferh_magnetisation_curve}
\end{figure}

The three interaction parameters $J_{\langle 001 \rangle}$, $J_{\langle 011 \rangle}$ and $D_{\langle Q \rangle}$ can be identified by agreement with experimental magnetization data for the FM phase, where $T_{\mathrm{M}}$ and the Curie temperature $T_{\mathrm{C}}$ determine their ratios and magnitudes. The values $J_{\langle 001 \rangle} = 0.40\times\E{-21}$J, $J_{\langle 011 \rangle} = 2.75\times\E{-21}$J and $D_{\langle Q \rangle} = 0.23\times\E{-21}$J give a good agreement with experiments (Fig.~\ref{fig:ferh_magnetisation_curve}a). It is also remarkable that the shape of the temperature dependent magnetization is similar to the Heisenberg model, a factor which is related to the material behaving as a classical, rather than quantum magnetic system.

At $T_{\mathrm{M}}$ we observe a significantly larger thermal hysteresis than the experimental results. This is primarily a result of our finite system size of 32$\times$32$\times$32 (32768 spins) with periodic boundaries. The lack of critical behavior (for example the divergence of the spin-spin correlation function) at a first order phase transition means that finite size effects can be complicated and the scaling behavior is usually as associated with the interface energy between the mixed phase, where in this case both AFM and FM order coexist at the same temperature. A limited system size restricts the maximum extent of nucleated domains of each ordering and hence the size of the interface region. Experimental observations of small FeRh nanoparticles~\cite{Ko:2008ke} and thin films~\cite{Han:2013wb} have also shown a large increase in the thermal hysteresis and even the suppression of the AFM phase~\cite{Hillion:2013wq}.

Fig.~\ref{fig:ferh_magnetisation_curve}b shows the contribution to the total energy of each Hamiltonian term as a function of the temperature. The four spin term has a greater thermal scaling exponent than the bilinear terms. Below $T_{\mathrm{M}}$, the bilinear terms scale as $[J_{\langle001
\rangle}(T)/J_{\langle001
\rangle}(0)] = [M(T)/M(0)]^{1.93}$ and $[J_{\langle001
\rangle}(T)/J_{\langle001
\rangle}(0)] = [M(T)/M(0)]^{1.55}$ whereas the four spin term scales with the much higher exponent $[D_{\langle Q\rangle}(T)/D_{\langle Q\rangle}(0)] = [M(T)/M(0)]^{3.48}$. At $T_{\mathrm{M}}$ a crossover behavior occurs and the ordered state of the system is now determined by the bilinear interactions.

\begin{figure*}
  \includegraphics[width=0.98\textwidth]{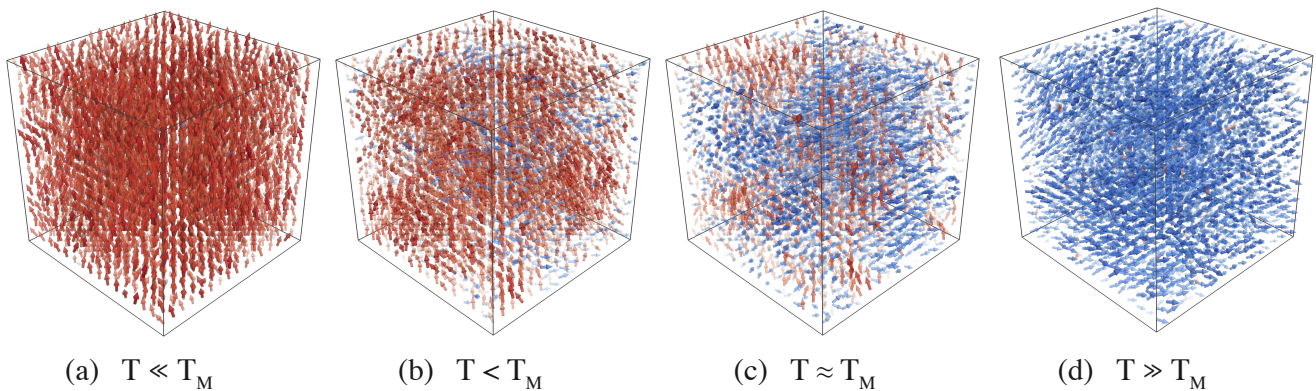}
  \caption{(Color online)
    Visualization of the Neel vector (red) and magnetization (blue) averaged over each unit cell as the model is heated through $T_{\mathrm{M}}$. (a) Below $T_{\mathrm{M}}$ the system is purely AFM (b) approaching $T_{\mathrm{M}}$ small regions of FM order nucleate (c) at $T_{\mathrm{M}}$ a mixed phase exists where AFM and FM domains coexist (d) raising the temperature above $T_{\mathrm{M}}$ leads to a purely FM phase.
  }
\label{fig:ferh_visualisation}
\end{figure*}

A characteristic aspect of first order phase transitions is the existence of a mixed phase~\cite{Goldenfeld:1992ti}. At the point of the phase transition there are two equal free energy minima, one associated with each phase and so AFM and FM order coexist. Experimentally, the mixed phase is hard to study directly at short time scales because the FM domains are initially randomly oriented~\cite{Bergman:2006iq}. On larger length and time scales a FM domain structure emerges~\cite{Baldasseroni:2012dy}. Simulation allows the direct observation of the mixed phase (Fig.~\ref{fig:ferh_visualisation}). Approaching the phase transition ($T < T_{\mathrm{M}}$), small areas of FM order nucleate with random orientation. At the transition temperature ($T \approx T_{\mathrm{M}}$), these regions combine and the lattice contains both AFM and FM regions, each with a common orientation. The latent heat of the system must be overcome for the phase transition to be complete, leaving only FM order. The length scale represented here is small ($\approx 10$nm) compared to thin film experiments where multiple independent regions would be expected to develop.

\begin{figure}
  \includegraphics[width=0.48\textwidth]{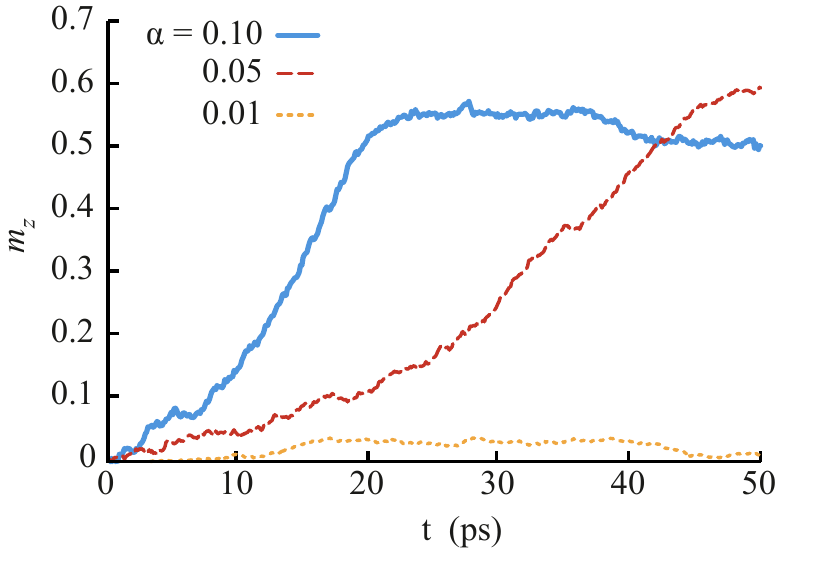}
  \caption{(Color Online) The $z$-component of the magnetization after laser heating using the two-temperature model to generating temperature profiles for a $100$fs laser pulse. Different Gilbert damping parameters are used, showing a strong dependence on the characteristic time scales.}
\label{fig:ferh_2tm}
\end{figure}

The results of pump-probe laser heating experiments have shown that FM phase can begin nucleating within the first picosecond after the application of a 100-150~fs laser pulse~\cite{Ju:2004vp,Thiele:2004wv}. The observed magnetic response is faster than that of the lattice. The dynamical approach of ASD allows the study of the time scale in which FM order is formed after laser heating. We use the so-called two-temperature model (2TM)~\cite{Kaganov:1957wf,Chen:2006bo} to represent the laser heating, whereby the laser power is assumed to be deposited into the electronic system which rapidly increases in temperature within a few picoseconds. On a timescale of tens of picoseconds the phonon temperature equilibrates with the electronic temperature. The spin system is generally coupled to the electronic temperature, although is not necessarily in equilibrium with it. The strength with which energy is exchanged into the spin system is mostly determined by the Gilbert damping parameter. There have been two attempts to identify this parameter experimentally in FeRh. Bergman et al.~\cite{Bergman:2006iq} inferred the value from a FM domain growth model, resulting in the unusually large value of $\alpha=0.3$, although they note that this large value includes extra effects beyond the intrinsic material value. Mancini et al.~\cite{Mancini:2013ut} attempted to directly measure the material value of $\alpha$ and came to a value of $\alpha=0.0013\pm0.0008$ in the FM phase, however they find that this value increases approaching the phase transition. The value of the Gilbert damping in the AFM phase is completely unknown and the energy transfer during laser heating may not be well represented by static Gilbert damping measurements.

We perform simulations with laser heating using three values of the Gilbert damping, $\alpha=0.10$, $0.05$ and $0.01$ (Fig.~\ref{fig:ferh_2tm}). The pulse width is $100$fs and the 2TM parameters are those of a typical metal, the same as used in Ref.\onlinecite{Vahaplar:2009fp}. For the lowest damping $\alpha=0.01$ the coupling is insufficiently strong between the spin system and the lattice for the laser pulse to heat the spin system across $T_{\mathrm{M}}$ and the associated latent heat, before the electron temperature equilibrates with the phonon temperature. For the values of $\alpha=0.10$ and $0.05$ the FM phase is formed and stabilized, although it is only for the higher value of $\alpha=0.10$ where significant FM order appears on a picosecond time scale. The time scale taken to reach magnetic saturation is $\approx 20$~ps which is also in agreement with experiment. If one wishes to make a complete comparison with experiment then there is a large parameter space to be searched, including the parameters of the 2TM. However this result shows that in principle the FM order can be generated on this short time scale, emanating only from the competition between the magnetic interactions. Increasing values of $\alpha$ cause the formation of FM order on a shorter timescale and it would greatly benefit the understanding of the phase transition in FeRh if the intrinsic damping were known in the AFM phase and the transition region. The difference in character between FM and AFM spin waves as well as the inertial dynamics of antiferromagnets~\cite{Kimel:2009ca} may also play an important role in this time scale.

We have shown that the metamagnetic phase transition in FeRh can be explained in terms of a competition between bilinear and higher order four spin effective exchange interactions which occur due to the non-linear dependence of the Rh moment on the orientation of the surrounding Fe moments. The model was parameterized from experiment and found to give a first order phase transition, showing important characteristics such as the mixed phase, which can be hard to resolve experimentally on the small scale. Importantly, our use of ASD allows identification the time scale associated with the phase transition. We show that a purely magnetic mechanism for the phase transition is viable even at the picosecond time scale, i.e. faster than magnetovolume effects. A likely explanation for the volume expansion is the latent heat of the first order phase transition, although this is yet to be investigated.

\begin{acknowledgments}
    The authors would like to thank useful discussion with O.N.~Mryasov and L.~Szunyogh. The work was supported by the EU Seventh Framework Programme under grant agreement No.~281043, FEMTOSPIN
\end{acknowledgments}

\end{document}